\documentclass[12pt]{article}
\usepackage{amssymb,amsmath,amsfonts,amsthm,
  natbib,setspace,dsfont,enumerate,anysize,sectsty,
  graphicx,algorithm,algorithmic,caption}

\usepackage{mathptmx}
\usepackage{txfonts}
\allsectionsfont{\normalfont\sffamily\bfseries}

\usepackage[svgnames]{xcolor}

\marginsize{1in}{.9in}{.5in}{1.1in}

\usepackage{fancyhdr}
\usepackage[utf8]{inputenc} 
\usepackage[T1]{fontenc}    
\usepackage{comment}       
\usepackage{microtype}

\DeclareFontFamily{OT1}{pzc}{}
\DeclareFontShape{OT1}{pzc}{m}{it}{<-> s * [1.200] pzcmi7t}{}
\DeclareMathAlphabet{\mathscr}{OT1}{pzc}{m}{it}

\newcommand{\mc}[1]{\mathcal{#1}}
\newcommand{\mr}[1]{\mathrm{#1}}

\newcommand{\ds}[1]{\mathds{#1}}

\newcommand{\indep}{\perp\!\!\!\!\!\perp}

\newcommand{\outcome}{\ensuremath{y}}
\newcommand{\struct}{\ensuremath{~\mathscr{g}}}
\newcommand{\cf}{\ensuremath{\mathscr{h}}}
\newcommand{\cfEst}{\ensuremath{\skew{4}\hat{\mathscr{h}}}}
\newcommand{\price}{\ensuremath{p}}

\DeclareMathOperator*{\E}{\mathds{E}}

\usepackage{todonotes}

\begin{document}
\setstretch{1.5}\pagestyle{empty}	


\vskip 1cm
\noindent
{\Huge\sf
Counterfactual Prediction with 

\vskip .2cm
\noindent 
Deep Instrumental Variables Networks
}

\vskip 1cm
{\sf\large 

\noindent
Jason Hartford, \textit{University of British Columbia}

\noindent
Greg Lewis, \textit{Microsoft Research and NBER}

\noindent
Kevin Leyton-Brown, \textit{University of British Columbia}

\noindent
Matt Taddy, \textit{Microsoft  Research and the University of Chicago}

}

\vskip 2cm
\setstretch{1.1}
{\noindent \small  
We are in the middle of a remarkable rise in the use and capability of artificial intelligence.  Much of this growth has been fueled by the success of deep learning architectures: models that map from observables to outputs via multiple layers of latent representations.  These deep learning algorithms are effective tools for unstructured prediction, and they can be combined in AI systems to solve complex automated reasoning problems. This paper provides a recipe for combining  ML algorithms to solve for causal  effects in the presence of instrumental variables -- sources of treatment randomization that are conditionally independent from the response.  We show that a  flexible IV specification resolves into two prediction tasks that can be solved with deep neural nets: a first-stage network for treatment prediction and a second-stage network whose  loss function involves integration over the conditional treatment distribution.  This {\it Deep IV} framework imposes  some specific structure on the stochastic gradient descent routine used for  training, but it is general enough that we can take advantage of off-the-shelf ML capabilities and avoid extensive algorithm customization.  We outline how to obtain out-of-sample
causal validation in order to avoid over-fit.  We also introduce schemes for both Bayesian and frequentist inference: the former via a novel adaptation of dropout training, and the latter via a data splitting routine.  
}
	
\newpage
\setstretch{1.4}
\pagestyle{plain}

\section{Introduction}

Supervised machine learning (ML) provides a myriad of effective methods for
solving prediction tasks.  
In these tasks, the learning algorithm is trained and
validated to do a good job predicting the outcome for future examples from the
same data generating process (DGP).  However, decision makers (and automated
decision systems)  look to the data to model the effects of a {\it policy
change}.   Precisely because  policy is going to change,  the future
relationship between inputs and outcomes will be different from what is in the
training data.  The ML algorithm will do a poor job of predicting the many
potential futures associated with each policy option.
 
For example, optimal pricing  requires predicting sales under {\it changes} to
prices, a doctor needs to know how a patient will respond to various {\it treatment}
options, and advertisers want to 
identify ads that {\it cause} sales.  In order to accurately answer such 
{\it counterfactual} questions it is necessary to model the structural (or causal)
relationship between policy (i.e., treatment) and outcome variables.  Randomized control
(`AB') trials are the gold standard for establishing causal relationships, but
conducting them is often impractical or excessively expensive. Observational
data, by contrast, is abundant. This paper uses the concept of instrumental variables to construct systems of machine learning (ML) tasks that can be applied in causal inference.  We develop recipes for application of deep neural nets (DNNs) within these systems, with the result that we are able to leverage supervised  ML to establish
causal relationships in large and unstructured  datasets.

The instrumental variables (IV) framework is a general class of methods for using observational data to establish causal
relationships.  It has a long history, especially in economics   \citep[e.g.,][]{wright28,reiersol45}. The idea is to use sets of variables that only affect treatment
assignment and not the outcome variable---so-called \emph{instruments}---to consistently estimate the causal treatment effect.  
The framework is most straightforward in the case of an imperfect experiment.  Consider a scenario where one of the inputs to treatment assignment has been  randomized, but where other influences  are potentially  endogenous: they are connected to unobserved influences on the outcome.  For example, in a medical trial we might have a treatment that is made available to a random sample of patients. However, only a portion of those patients actually take the treatment (perhaps because it causes discomfort).  In this scenario, the random availability of treatment is our instrument and an IV analysis is used to infer the causal treatment effect in the face of  selective partial adherence.

\begin{figure}
\centering
~~~~\includegraphics[width=.3\textwidth]{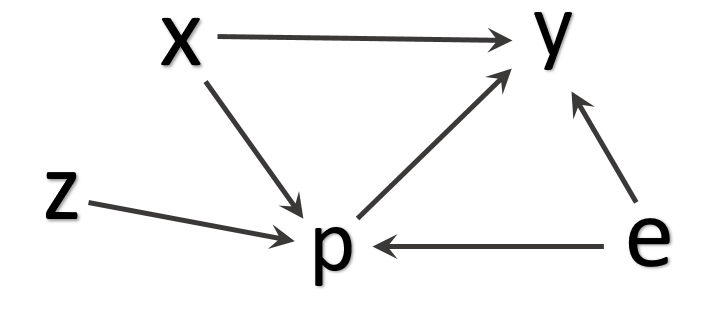}~~~~~~
\includegraphics[width=.6\textwidth]{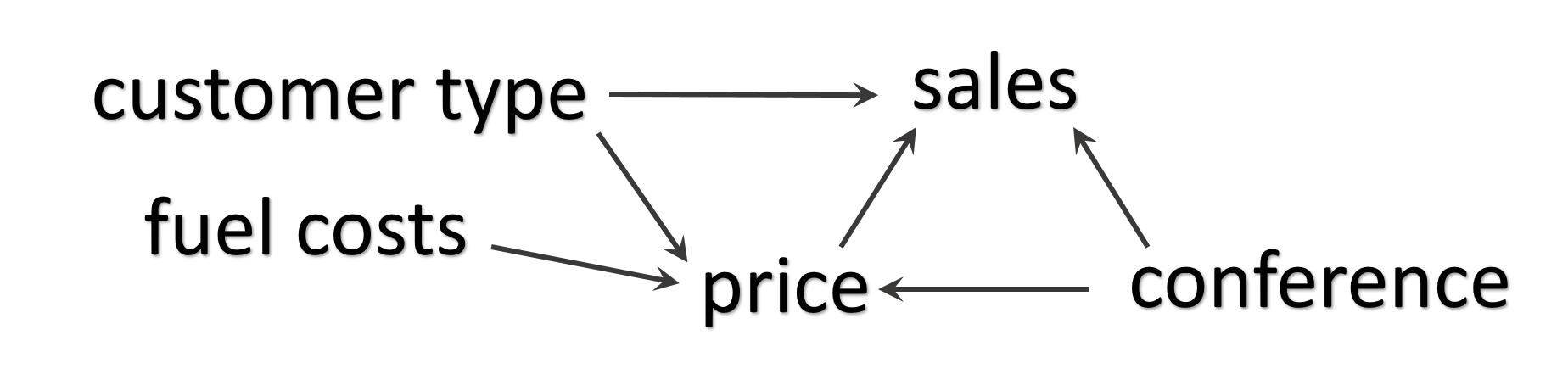}
\caption{\label{fig:ivgraph}  A graphical model showing a simple version of the structure of the observational DGP under our IV specification.  In addition, we will assume that $e$ influences $y$ as an {\it additive} effect.  On the right, we give names to the variables in the context of our simple air-travel demand example.  Price is the policy variable, sales is the outcome, and there are (say) business- and leisure-travel customer types as observable covariates.  There is a big `conference', unobserved to the policy-maker, that drives demand and (due to the airline's pricing algorithms) price.  The instrument is the cost of fuel, which  influences sales only via price.}
\end{figure}

The full scope of IV analysis comes from its use in so-called {\it natural experiments}, where the instruments are contained in observational data and are not the result of intentional randomization.  For example, a policy-maker might want to understand consumer price sensitivity to airfare (e.g., in setting an `airport improvement fee'): they need to model the effect of price (treatment) on sales (outcome). Prices vary for many reasons, some of them driven by demand. Regression of price on sales will fail to capture the true causal relationship: prices are high around holidays {\it because} demand is high and naive analysis will say that higher prices lead to more sales (nonsense).  The problem is existence of factors that lead to co-movement in price and demand. Some of these are observable, such as major holidays, but many will be unknown to the policy-maker (on-line search activity, route capacity, etc).  
However, you could argue that cost of fuel is an {\it instrument}: it varies for reasons independent of demand and affects sales only via ticket prices. We can thus understand the causal effect of price on sales by observing how demand varies with the cost of fuel.  Changes in fuel cost create movement in ticket prices that is {\it independent} of unobserved demand shifts, and this movement is as good as randomization for the purposes of causal inference.  See Figure \ref{fig:ivgraph} for a graphical illustration of this scenario.

Most  IV applications make use of a two-stage least squares procedure \citep[2SLS; e.g.,][]{angrist1996identification} that requires  assumptions of linearity and homogeneity (e.g., all airline customers must have the same price sensitivity).  Nonparametric IV methods from the econometrics literature relax these assumptions \cite[e.g.,][]{newey2003instrumental}.  However, these methods work by modeling the outcome as an unknown linear combination of a pre-specified set of basis functions of the treatment and other covariates (e.g. Hermite polynomials, wavelets or splines) {\it and} then modeling the conditional expectation of each of these basis functions in terms of the instruments (i.e. the number of equations is quadratic in the number of basis functions). This requires a strong prior understanding of the DGP by the researcher, and the complexity of both specification and estimation explodes when there are more than a handful of inputs. 

Advances in deep learning have demonstrated the effectiveness for prediction tasks of learning latent representations of complex features spaces \citep[for recent surveys, see][]{deepsurvey,schmidhuber2015}.  We want to use these powerful learning algorithms for IV analysis. We do this by deriving from the IV structure a {\it system} of machine learning tasks that can each be targeted with deep learning and which, when solved, allow us to make counterfactual claims and perform causal inference.  We break IV analysis into two ML stages: a first stage that models the conditional distribution for treatment given the instruments and covariates, and a second stage which targets a loss function involving integration over the conditional treatment distribution from the first stage.  Both stages are use deep neural nets trained via stochastic gradient descent \citep{robbins1951stochastic,bottou2010large} and out-of-sample causal validation  guards against over-fit.  We refer to this setup as the Deep IV framework.

\noindent
We highlight the main contributions to the ML and AI literatures.

\vskip .1cm
\noindent
{\bf Economic AI:}  This work uses econometric theory to resolve economic measurement questions into prediction tasks that can be targeted with deep learning algorithms.  It is among the first examples of such a strategy, building from an earlier literature that combines ML with econometrics, e.g., for sparse linear models in \cite{belloni_sparse_2012,belloni2014inference} and for trees and forests in \cite{athey2016recursive,wager2015inference}.\footnote{In addition, the recent paper by \cite{chernozhukov2016double} describes use of generic machine learning in applications where you have a randomized experiment conditional upon the controls (i.e., conditional ignorability).  }  Our approach is designed to take advantage of {\it generic} ML capabilities and avoid extensive algorithm customization.  This allows state-of-the-art DNN architectures to be plugged into economically relevant systems, and we hope that it will lead more ML researchers to target their efforts towards solving problems in this important domain.  In our specific applications, this strategy lets us infer user- and time-specific treatment effects from complicated data involving raw text, images, and detailed hierarchical information.

\vskip .1cm
\noindent
{\bf Causal Validation:} In a standard predictive ML setting, out-of-sample (OOS) validation is essential for setting tuning parameters and avoiding over-fit.  Simple schemes, such as cross validation, do not work for causal inference because the available data is not immediately representative of the potential counterfactual outcomes.  However, our framework resolves causal inference into two prediction problems and we show how our fit for each of these tasks can be evaluated on left-out data.  This provides a general approach to model tuning and validation for causal inference.

\vskip .1cm
\noindent
{\bf Integral SGD:} The second stage of our Deep IV framework involves training a DNN whose role in the loss function is integrated over a distribution of possible inputs.  In our case, the integral is with respect to the conditional treatment distribution; however, similar setups show up elsewhere in ML \citep[e.g., the variational autoencoders of][]{kingma2013auto}.
We specify how SGD should be applied when these integrals are high-dimensional and need to be solved via Monte Carlo (MC) sampling: {\it independent} samples should be used for each instance of the integral in a gradient calculation, and it is most efficient to use a single MC draw for each gradient observation.  These results will be generally applicable for learning algorithms that entail integral loss functions.

\vskip .1cm
\noindent
{\bf Inference:}  We provide techniques for both Bayesian and frequentist inference about the treatment effects.  In the first case, we show that a network trained under dropout \citep{srivastava2014dropout} can be interpreted as a very simple form of variational Bayesian inference with a posterior precision that is fixed by the dropout rate.  This provides a more direct and interpretable connection between dropout and Bayesian inference than that derived in \cite{gal2015dropout} (it also offers guidance on how to choose the dropout rate).  In the second case, we use a sample-splitting strategy that allows us to study the fitted network structure, and its implications for treatment effects, via two-stage least-squares inference with left-out data. 

\vskip .25cm

Section \ref{sec:generalframework} describes our general IV specification and its resolution into two  ML tasks, while Section \ref{sec:estim} outlines neural network estimation for these tasks with particular attention on the SGD routine used in model training.  Section \ref{sec:inference} then details both Bayesian and frequentist inference results and Section \ref{sec:sim} provides simulation experiments to  illustrate our methods.

\section{Machine learning  for counterfactual prediction}
\label{sec:generalframework}

Consider the following {\it structural} equation with additive latent errors,
\begin{equation} 
\label{additive}
\outcome = \struct (\price,x) + e,
\end{equation}
where $\outcome$ is the outcome variable (e.g., sales in our airline example), $\price$ is the policy or treatment variable (e.g., price), and $x$ is a vector of observable covariate features (e.g., time and customer characteristics).  The function $\struct(\cdot)$ is some unknown and potentially non-linear continuous function of $x$ and $\price$, and the `unobservable' error $e$ enters additively with unconditional mean $\E e = 0$. We emphasize that, in contrast to the usual machine learning setup, here the errors are potentially correlated with the inputs: $\ds{E}[e|x,p] \neq 0$ and, in particular, $\ds{E}[pe|x] \neq 0$.  
That is to say, the policy variable $p$ is \emph{endogenous}. 

Define the counterfactual prediction function 
\begin{equation}
\cf(\price,x) \equiv \struct (\price,x) + \E[e|x],
\end{equation}
 which is the conditional expectation of $\outcome$ given the observables $\price$ and $x$, \emph{holding the distribution of $e$ constant as $\price$ is changed}.
Note that we condition only on $x$ and not $\price$ in the term $E[e|x]$; this term is typically nonzero, but it will remain constant under arbitrary policy changes.\footnote{It may be easier to think about a setting where $e\indep x$, so that the latent error is simply {\it defined} as being due to factors orthogonal to the observable controls.  In that case, $\cf(\price,x) = \struct (\price,x)$.  All of our results apply in either setup. }  Thus $\cf(\price,x)$ is the structural estimand of interest;
 to evaluate policy options, say $\price_0$ and $\price_1$, we can look at the difference in mean outcomes $\cf(\price_1,x)-\cf(\price_0,x) = \struct(\price_1,x)-\struct(\price_0,x)$.

In standard `unstructured' ML, the prediction model is trained to 
fit $\E[y|\price, x]$.  This estimand will typically be {\it biased} against our structural objective: 
\begin{equation}\label{eq:unsctruct}
\E[y|\price, x] = \struct (\price,x) + \E[e|\price,x] \neq \cf(\price,x)
\end{equation}
since presence of error endogeneity -- $\ds{E}[pe|x] \neq 0$ -- implies that $\E[e|\price,x] \neq \E[e|x]$.  Endogeneity occurs whenever unobservables are correlated with both policy and  outcome, as was the case for price and sales in our demand example.  This object is inappropriate for policy analysis as it will lead to biased counterfactuals: 
\begin{equation}
\E[y|\price_1, x] - \E[y|\price_0, x] = \struct (\price_1,x) - \struct (\price_0,x) + \left(\E[e|\price_1,x]-\E[e|\price_0,x]\right).
\end{equation}

Fortunately, the presence of \emph{instruments} allows us to estimate an unbiased $\cfEst(\price,x)$ that captures the structural relationship between $\price$ and $\outcome$.  These are variables, say $z$, that satisfy three conditions:
\begin{itemize}
	\item[] \emph{Relevance:} $F(\price|x,z)$, the distribution of $\price$ given $x$ and $z$, is not constant in $z$,
	\item[] \emph{Exclusion:} z does not enter equation~(\ref{additive}) -- i.e., $z \indep y \; | \; (x,p,e)$,
	\item[] \emph{Unconfounded Instrument:}  $z$ is conditionally independent of the error -- i.e., $z \indep e \; | \; x$.\footnote{Under the additive error assumption made in~(\ref{additive}), unconfoundedness of the instrument is not necessary: we could replace this assumption with the weaker mean independence assumption $E[e|x,z] = 0$ without changing anything that follows.  We use the stronger assumption to facilitate extensions, e.g. to estimating counterfactual quantiles.} 
\end{itemize}
The last condition is chosen to match the `unconfoundedness' assumption in the Neyman-Rubin potential outcomes framework \citep{rosenbaumrubin1983} (i.e. $p \indep e \; | \; x$). 
But our assumption is weaker --- in particular, we allow for $p \not \indep e | x$ --- and so the matching and propensity-score re-weighting approaches often used in that literature will not work here.    
Figure \ref{fig:ivgraph} presents a graphical model summarizing a simple IV system.

Taking the  expectation of both sides of (\ref{additive}) conditional on $[x,z]$ and applying these assumptions establishes the relationship \citep[cf.][]{newey2003instrumental}:
\begin{align}\label{sequence} \E[\outcome|x,z] 
 ~=~ \E[\struct (\price,x)|x,z] + \E[e|x] ~=~ \int \cf(\price,x) dF(\price|x,z)  \end{align}
where, again, $dF(\price|x,z)$ is the conditional treatment distribution.  This shows that we can reason about our counterfactual prediction function $\cf(\price,x)$ (the RHS of~(\ref{sequence})) by learning to predict $\E[\outcome|x,z]$ (the LHS of~(\ref{sequence})). The relationship in (\ref{sequence}) defines an {\it inverse problem} for $\cf$ in terms of two directly observable  functions: $\ds{E}[y|x,z]$ and $F(p|X,z)$.  Specifically, to minimize $\ell_2$ loss given $T$ data points and given a function space $\mathcal{H}$ (which may not include the true \cf), we solve
\begin{equation}\label{constrained} \min_{\cfEst \in \mathcal{H}} \sum_{t=1}^T\left(y_t - \int \cfEst(\price,x) dF(\price|x_t,z_t)\right)^2. \end{equation}
Since the  treatment distribution is unknown,  IV analysis typically splits into two stages: a first to estimate $\hat F(\price|x_t,z_t) \approx F(\price|x_t,z_t)$, and a second to minimize (\ref{constrained}) after replacing $F$ with $\hat F$.

Existing approaches to IV analysis rely upon a {\it linearization} of both $\cfEst$ and $\hat F$.  For example, the popular two-stage least-squares (2SLS) procedure posits
\begin{align}\label{2sls}
\outcome &= \gamma \price + x\beta_\outcome + e, \\
\price &= \tau z + x\beta_\price + v \notag
\end{align}
with the assumptions that $E[e|x,z] = 0$, $E[v|x,z] = 0$ and $E[ev] \neq 0$ (which implies $E[ep] \neq 0$).   
In this case, the integral in (\ref{constrained}) simplifies as 
$$
\int \cfEst(\price,x) dF(\price|x,z) = \gamma\E[\price|x,z] + x\beta_\outcome + \E[e|x] = \gamma(\tau z + x\beta_\price)  + x\beta_\outcome,
 $$
The 2SLS approach is to replace the unknown $\E[\price|x,z]$ with an estimate $\hat \price = \hat\tau z + x\hat\beta_\price$.
In the `first-stage' $\hat \price$ is estimated by an ordinary least squares (OLS) regression of $p$ on $x$ and $z$.  You then run a `second-stage' regression of $y = \gamma \hat \price + x \beta_y$, again by OLS.  


The 2SLS procedure is a straightforward and statistically efficient way to estimate the effect of the policy variable (i.e. $\gamma$), but requires two strong assumptions: linearity (i.e., that both first and second stage regressions
are correctly specified) and homegeneity (i.e., that the policy affects all
individuals in the same way).\footnote{The estimated $\hat \gamma$ remains interpretable as a `local average treatment effect' (LATE) under less stringent assumptions \citep[see][for an overview]{angrist1996identification}.}  
Flexible nonparametric extensions of 2SLS, such as in \cite{newey2003instrumental},
replace the simple linear regressions in (\ref{2sls}) with a linear projection onto a series of known basis functions. The outcome equation is replaced with $\outcome =
\sum_k \gamma_k\varphi_k(\price,x) + e$, where $\varphi_k$ are pre-specified
functions and the conditional expectation $\E[\varphi_k(\price,x)|x,z]$ of {\it each} of these bases is estimated in a series of non-parametric first stage regressions.  
The non-parametric first stage estimators are often series estimators, and so require another basis expansion (now in $[x,z]$).\footnote{The model in \cite{newey2003instrumental} has $\widehat\E[\varphi_k(\price,x)|x,z] = \sum \alpha_{jk} \beta_j(x,z)$, and they apply OLS to obtain $\hat\alpha_{jk}$ for the pre-specified first-stage basis functions $\beta_j$. Alternatively, \cite{BCK2007} write the first-stage basis expansion as  $\widehat\E[y - \sum_k \gamma_k \varphi_k(\price,x)|x,z] = \sum \alpha_{j} \beta_j(x,z)$, which reduces the number of series coefficients to estimate but leads to a nested optimization setup. This is the same setup studied, in more generality, in \cite{chenpouzo12}.  In addition, see \cite{hall2005} and \cite{darolles2011nonparametric} for approaches based on kernel methods. } Thus, although this is an
effective strategy for introducing flexibility and heterogeneity with low
dimensional inputs, this system of series estimators becomes computationally intractable
for anything beyond trivially low-dimensional $[x,z]$.

This paper proposes to avoid explicit linearization by instead directly
targeting the integral objective in (\ref{constrained}).  In the case where
$\price$ is low dimensional, we replace the integral with a sum over its
support.  When this is not feasible, we use efficient Monte Carlo integration
inside a stochastic gradient descent routine.  In either case,  $F$ and
$\cf$ can be learned by  any generic ML model that can be trained via
gradient descent. We make use of deep neural networks, allowing us to take
advantage of  state-of-the-art  ML technology, and provide a recipe for 
two-stage deep learning. The next section details these ideas and their
implementation.

\section{Estimation and neural network implementation}
\label{sec:estim}

As mentioned, our approach proceeds in two stages.  We outline each in-turn  before discussing Monte Carlo SGD for our second-stage objective and a framework for out-of-sample validation.

\subsection{First stage: Treatment network}

In the first stage we learn $F(\price|x, z)$ using an appropriately chosen distribution parameterized by a DNN, say $\hat F = F_\phi(\price|x, z)$ where $\phi$ is the set of network parameters. Since we will be integrating over  $F_\phi$ in the second stage, we must fully specify this distribution. Estimation then proceeds by maximum likelihood via SGD on the implied negative log likelihood.

In the case of discrete $\price$, we model $F_\phi(\price|x,z)$ as a
multinomial  $\mathrm{MN}\left(\price\mid \pi(x,z; \phi)\right)$
with $\mr{p}(\price=\price^k) = \pi_k(x,z;\phi)$ for each treatment category $\price^k$ and where
$\pi_k(x,z;\phi)$ is given by a DNN with softmax output. For continuous
treatment, we model $F$ as a mixture of Gaussian distributions  $$ F_\phi
(\price|x,z) = \sum_k \pi_k(x,z;\phi) \mathrm{N}\left(\mu_k(x,z;\phi),
\sigma^2_k(x,z;\phi)\right),$$  where component weights $\pi_k(x,z;\theta)$ and
parameters $\left[\mu_k(x,z;\phi), \sigma_k(x,z;\phi)\right]$ form the final
layer of a neural network parametrized by $\phi$.  This model is known as a
mixture density network, as detailed in \S 5.6 of  \cite{bishop_pattern_2006}.
With enough mixture components it can approximate arbitrary smooth
densities.  Mixed continuous-discrete distributions are obtained by
replacing some mixture components with a point mass.  In each case, fitting
$F_\phi$ is a standard ML task.

\subsection{Second stage: Outcome network} 

In the second stage, our counterfactual prediction function $\cf$ is approximated by a DNN with real-valued output, say $\cf_\theta$.
Following from (\ref{constrained}), network parameters $\theta$ are optimized to minimize the integral loss function, over training data $D$ of size $T = |D|$ from the joint DGP $\mc{D}$,
\begin{equation}
\label{cfloss} 
\mathcal{L}(D; \theta) =  |D|^{-1}\sum_t\left(y_t - \int \cf_\theta(\price,x_t) dF_\phi(\price|x_t,z_t)\right)^2.
\end{equation}
Note that this loss involves the {\it estimated} treatment distribution function, $F_\phi$, from our first stage.\footnote{
We can replace (\ref{cfloss}) with other functions, e.g. a logit for categorical outcomes, but will use $\ell_2$ loss for most of our exposition.  Also, note that Bayesian inference introduces a second integral over the posterior distribution on $\phi$.}  

We use stochastic gradient descent \citep[SGD; see specific algorithms in, e.g.,][]{duchi2011adaptive,kingma2015adam} to train the network weights.  For $F_\phi$, standard off-the-shelf methods apply.  Our second stage optimization, for $\cf_\theta$, needs to account for the integral in (\ref{cfloss}).  The noisy SGD sample gradient, for a single observation $d_t = [x_t,z_t,\price_t,\outcome_t]$,  is avai{}lable as
\begin{equation}
\label{genericgrad}
\nabla_\theta \mathcal{L}_t \equiv \frac{\partial\mathcal{L}(d_t; \theta)}{\partial \theta} = 
 \left(y_t - \int \cf_\theta(\price,x_t) dF_\phi(\price|x_t,z_t)\right) 
 \int \cf_\theta'(\price,x_t)dF_\phi(\price|x_t,z_t),
\end{equation}
assuming continuity of both $\cf_\theta$ and $\cf_\theta'$,  where
$\cf_\theta' = \partial\cf_\theta/\partial\theta$. When the policy space is discrete and low dimensional, in which case Section \ref{sec:estim}.1 specifies a multinomial response network for $F_\phi$, this gradient can be expressed exactly as
\begin{equation}
\label{discretegrad}
\nabla_\theta \mathcal{L}_t = 
 -2\left(y_t - \sum_k \pi_k(x_t,z_t;\phi) \cf_\theta(\price^k,x_t) \right) 
 \sum_k \pi_k(x_t,z_t;\phi)\cf_\theta'(\price^k,x_t).
\end{equation}
For more complicated and continuous treatments, the integrals must be approximated.  The next section proposes an efficient Monte Carlo approximation SGD algorithm.

\subsection{SGD with Monte Carlo integration}

SGD convergence requires that each sampled gradient $\nabla_\theta \mathcal{L}_t$ is {\it unbiased} for the population gradient, $\nabla_\theta \mathcal{L}(\mc{D}; \theta)$.  Lower variance for $\nabla_\theta \mathcal{L}_t$ will tend to yield faster convergence \citep{zinkevich2003online} while the computational efficiency of SGD on large datasets  requires limiting the number of operations going into each gradient calculation \citep{bousquet2008tradeoffs}.  We are thus free to replace the integrals in (\ref{genericgrad}) with anything that leads to unbiased estimates of the complete gradient, and should do so to minimize variance under a constraint on the number of operations involved.

Basic Monte Carlo (MC) methods replace an integral with respect to a probability measure with the average of draws from the associated probability distribution:  $\int \mathscr{h}(p) dF(p) = \E_F\mathscr{h}(p) \approx B^{-1} \sum_b \mathscr{h}(p_b)$ for $\{p_b\}_1^B \stackrel{iid}{\sim} F$.  This method can be applied in our context with an important caveat: {\it independent} samples must be used for each instance of the integral in the gradient calculation.
To see this, note that (\ref{genericgrad}) has expectation
\begin{align}
\ds{E}_\mc{D}\nabla_\theta \mathcal{L}_t &= -2\ds{E}_\mc{D}\left( \ds{E}_{F_\phi(p|x_t,z_t)}\!\!\left[y_t - \cf_\theta(\price^k,x_t) \right] 
 \ds{E}_{F_\phi(p|x_t,z_t)}\!\!\left[ \cf_\theta'(\price^k,x_t) \right]\right) \label{equalgrad}\\
 & \neq -2\ds{E}_\mc{D} \ds{E}_{F_\phi(p|x_t,z_t)}\!\!\left[\left(y_t - \cf_\theta(\price^k,x_t)  
 \right)\cf_\theta'(\price^k,x_t) \right],\label{neqgrad}
\end{align}
where the inequality holds so long as $\mr{cov}_{F_\phi(p|x_t,z_t)}\!\!\left[\left(y_t - \cf_\theta(\price^k,x_t)  
 \right)\cf_\theta'(\price^k,x_t) \right] \neq 0$.  We thus need a gradient estimate based on unbiased MC estimates for each $\ds{E}_{F_\phi(p|x_t,z_t)}$ term in (\ref{equalgrad}).  This is obtained by taking two samples\footnote{These could be of different size, but for notational simplicity we keep them the same.} 
 $\{\dot{p}_b\}_1^B,  \{\ddot{p}_b\}_1^B \stackrel{iid}{\sim} F_\phi(p|x_t,z_t)$  and calculating the gradient  as
 \begin{equation}\label{mcgrad}
 \widehat \nabla_\theta^{B} \mathcal{L}_t \equiv -2\left(y_t - B^{-1}\sum_b \cf_\theta(\dot\price_b,x_t) \right) 
 B^{-1}\sum_b \cf_\theta'(\ddot\price_b,x_t).
 \end{equation}
Independence of the two samples ensures that $\ds{E}\widehat \nabla_\theta^{B} \mathcal{L}_t = 
\ds{E}_\mc{D}\nabla_\theta \mathcal{L}_t = \nabla_\theta \mathcal{L}(\mc{D};\theta)$, as desired.

The estimator in (\ref{mcgrad}) remains unbiased even in the case of $B=1$, where our gradient becomes
\begin{equation}\label{omemcgrad}
 \widehat \nabla_\theta \mathcal{L}_t \equiv \widehat \nabla_\theta^{1} \mathcal{L}_t
 = -2\left(y_t -  \cf_\theta(\dot\price,x_t) \right) 
\cf_\theta'(\ddot\price,x_t), ~~~\dot p, \ddot p \stackrel{iid}{\sim} F_\phi(p|x_t,z_t).
 \end{equation}
Indeed, this `two-draw' $\widehat \nabla_\theta \mathcal{L}_t$ is the optimal implementation of (\ref{mcgrad}) in a large-scale learning setting \citep{bousquet2008tradeoffs} where computation, rather than the amount of data available, is the binding constraint on error minimization.  Consider two alternative gradient estimators, both of which require roughly the same number of computational operations:  $\widehat \nabla_\theta^{M} \mathcal{L}_t$ and $M^{-1}\sum_{m=1}^M\widehat \nabla_\theta \mathcal{L}_{t_m}$, where this latter term is just the average of $M$ two-draw gradients across unique data points.
The variance of the first estimator is $M^{-1}\ds{E}_\mc{D}\left[\mr{var}_{F_\phi(p|x,z)}(\widehat \nabla_\theta \mathcal{L}_t)\right] + \mr{var}_\mc{D}\nabla_\theta\mathcal{L}_t $, which due to the second non-diminishing term will tend to be larger than $M^{-1}\mr{var}_\mc{D}\widehat \nabla_\theta \mathcal{L}_t$, the variance of the second estimator.  
Hence, efficient MC SGD will process more two-draw gradients $\widehat \nabla_\theta \mathcal{L}_t$ rather than spending time increasing the number of MC draws for a single sampled gradient.

As an aside, we note that this MC SGD algorithm -- and the requirement of independent sampling for each integral -- will apply to a wide range of loss functions that compose a norm and an integral.  This includes any implementation of the variational auto-encoders of \cite{kingma2013auto} that involve, e.g., $\ell_2$ observation loss.  In addition, there are a number of applications in economics and elsewhere that approach such loss functions through direct minimization of a monte carlo approximation to the true objective function (e.g., simulated maximum likelihood (SML) and simulated method of moments (SMM) \citep{mcfadden89, pakespollard89}). 
For example, in our case one would minimize the simulated objective
$\sum_t \left(y_t - B^{-1}\sum_b \cf_{\theta}(p^b_t,x_t)\right)^2$ for $\{p^b_t\}$ a sample of size $BT$ from $F_\phi(p|x_t,z_t)$, using an arbitrary minimization technique.
The simulated objective must be uniformly close (as a function of the parameters) to the original objective to guarantee that the solutions are close.
This requires $B$ large, which is computationally intensive and potentially prohibitive if $T$ is large.  
By contrast, our MC SGD scheme directly solves the original problem, and in a computationally efficient manner; it may thus be useful to economists in a large-data setting.  



\subsection{Causal validation}

Any complex ML framework requires an out-of-sample (OOS) validation procedure for tuning hyperparameters and optimization rules.  Fortunately,  Deep IV  can be validated with left-out data corresponding to each individual composite ML task.

Consider a left-out dataset $D_{lo}$.
 Our first stage, fitting $F_\phi$, is a standard density estimation problem and can be tuned to minimize the OOS deviance criterion
\begin{equation}\label{oos1}
\min_\phi \sum_{d_l\in D_{lo}} -\log f_\phi(p_l | x_l, z_l),   
\end{equation}  
where $f_\phi$ is either the probability mass function or density function associated with $F_\phi$, as appropriate.  
Second stage validation proceeds {\it conditional} upon a fitted $F_\phi$, and we seek to minimize the left-out loss criterion
\begin{equation}\label{oos2}
\min_\theta \sum_{d_l\in D_{lo}} 
\left(y_l - \int \cf_\theta(\price, x_l)dF_\phi(\price|x_l,z_l)\right)^2.
\end{equation}
The integral here can either be exact or MC approximate via sampling from $F_\phi$.  

Each stage is evaluated in turn, with second stage validation using the best-possible network as selected in the first stage.  Statistical `overfit' -- optimizing to noise -- will lead to a reduction in performance in each of these criteria.  In the first stage, using (\ref{oos1}) to avoid overfit guards against the `weak instruments' bias \citep{bound1995problems} that can occur when the instruments are only weakly correlated with the policy variable.  In the second stage, minimizing (\ref{oos2}) will lead to best-possible performance on the objective $\ds{E}[y|x,z]$ under the constraint imposed by $F_\phi$.  
Note that these criteria provide {\it relative} performance guidance: improving on each criterion will improve your performance on counterfactual prediction problems, but it does not tell you, e.g., how far $\cf_\theta(\price,x)$ is from true $\cf(\price,x)$.  That requires the inference steps of the next section.

\section{Inference}
\label{sec:inference}

In many ML applications, inference and uncertainty quantification  are of secondary importance after average predictive performance.  However, in policy decision settings it is crucial to know the credibility and variance of our counterfactual predictions.  This section addresses these needs.

\subsection{Frequentist inference and statistical properties}

Under the {\it relevance}, {\it exclusion}, and {\it unconfoundedness}  conditions of Section \ref{sec:generalframework}, which define the role of instruments,  nonparametric identification of the function $\cf$, known to be an element of some function space $\mc{H}$, simply requires that there is a unique solution for $\cf \in \mc{H}$ in   equation (\ref{sequence}): $\ds{E}[y|x,z] = \int \cf(p,x)dF(p|x,z)$.\footnote{This can be summarized via the `completeness condition' of \cite{newey2003instrumental}, which states that if  $\ds{E}[g(p,x)|x,z] = 0$ $\forall (x,z)$ in their support, then $g(p,x)=0$ $\forall (x,p)$.} 
When $p$ and $z$ are discrete, it is necessary and sufficient that the Markov matrix describing the conditional distribution of $p$ given $z$ for fixed $x$ has full rank with probability 1 (where the measure is over the distribution of $x$) \citep{newey2003instrumental}.
This says, roughly, that $z$ generates sufficient variation in $p$ for every $x$ that occurs with positive probability, which is necessary to identify heterogeneous treatment effects.

In practice, it is of primary importance to assess the validity of the structural IV conditions.  Outside of some special scenarios \citep{pearl1995testability}, exclusion and unconfoundedness are untestable and must  be {\it assumed}.  This is easy for randomized instruments (e.g.,  intent-to-treat), and there is extensive guidance in the economics literature for assessment of natural experiments \citep[see][for an introduction]{angrist2008mostly}.

Relevance can be observed: when instruments are non-informative at some $x$, then $F(p|z,x) = F(p|x)$. The  neural network will typically lead $F_\phi(p|z,x)$ to vary with $z$ when the instruments are unconditionally relevant, even at $x$ locations that see little instrument-treatment variation in the finite training sample.  That is, we will be semi-parametrically identified due to variation at neighboring covariate locations. We can also consider what happens when the first stage yields irrelevance at a given $x$: the second stage then targets $\left(y - \int \cf_\theta(p,x) dF_\phi(p|x)\right)^2$,  which at a {\it fixed} $x$ implies $\cf_\theta(p,x) = \cf_\theta(x) = \ds{E}[y|x]$, a constant.  
So where relevance fails the second stage objective pushes towards estimates of a zero policy effect.

If identification is satisfied, consistency of Deep IV follows from \citet[Lemma A1 and Theorem 4.1]{newey2003instrumental} if we treat both $F_\phi$ and $\cf_\theta$ as sieve estimators that grow in complexity so that they can arbitrarily closely approximate the true $F$ and $\cf$.\footnote{The universal approximation theorem of \cite{hornik1991approximation} implies that if the true $F_\phi$ and $\cf_\theta$ are sufficiently smooth, they can be approximated by neural nets of growing size. \cite{newey2003instrumental} require in addition that $\mc{H}$ is compact (to avoid discontinuities in solution to the ill-posed inverse problem). \cite{BCK2007} point out that compactness is fairly restrictive, but that bounded compactness achieves the same results and is more plausible in many settings.  \cite{chenpouzo12} relax these conditions and extend the theory to cover penalized sieve estimators.  We caution, however, that contemporary deep networks often involve a dramatic linear dimension reduction at the bottom layer; hence, sieve theory is a rough fit to common practice. }
Unfortunately, full Frequentist inference results for $\cf_\theta$, or for $\theta$, are not available via any reasonable asymptotic approximations.  Instead, we turn to a {\it data splitting} procedure wherein some portion of the data is left-out of the training sample and used to obtain {\it conditional} inference for $\cf_\theta$.  Splitting procedures like this have a long history \citep{cox1975note}, and have recently regained popularity as an option for  inference post model-selection \citep[e.g.,][]{fithian2015optimal,wager2015inference}.

Denote the $K_L$ nodes in the final layer of the response network (i.e., $\cf_\theta$) as 
$\eta_{k}(x,p)$, where the arguments $[x,p]$ are implictly fed through lower layers via weights $\theta$, and their conditional expectation at observation $i$ as
\begin{equation}\label{etabar}
 \bar\eta_{ik} = \ds{E}_{F_\phi( p|x_i,z_i)}\eta_k(x_i, p).
 \end{equation} 
As always, this expectation can be evaluated either exactly for discrete $p$ or via MC approximation.
Hence, the second stage network parameters, $\theta$, have been optimized to minimize
$
\sum_{d_i \in D} \left(y_i - \sum_k \bar\eta_{ik} \right)^2
$
over the in-sample training data $D$.  In addition, we introduce the shorthand 
$\eta_{ik} = \eta_k(x_i, p_i)$ for the node expressions at observed treatment values, and denote observation vectors as ${\eta}_i = [1~\eta_{i1} \cdots \eta_{iK_L}]'$ and ${\bar\eta}_i = [1~\bar\eta_{i1} \cdots \bar\eta_{iK_L}]'$.

Our data-splitting inference takes $\theta$ and $\phi$ as fixed after training and applies the fitted networks to calculate $\bar\eta_{lk}$ and $\eta_{lk}$ for {\it left-out} observations $d_l \in D_{lo}$.  
These values are then used as instruments and treatments, respectively, in standard linear IV estimation.
Moreover, assuming homoskedastic errors, the instruments are the ``right ones'': they are plug-in estimates of the infeasible optimal instruments, $E[\eta_{k}(x_i,p)|x_i,z_i]$ \citep{chamberlain1987}.

Letting $T_{lo} =|D_{lo}|$, stack the left-out `data' together as ${\bar H} = [{\bar\eta}_{1} \cdots {\bar\eta}_{T_{lo}}]'$,  ${H} = [{\eta}_{1} \cdots {\eta}_{T_{lo}}]'$, and $Y = [y_l \cdots y_{T_{lo}}]'$.
Then, since there are exactly as many instruments as treatments, the 2SLS estimate for the {\it causal} effect of ${\eta}_{l}$  on $y_l$ can be written in its method of moments  form as 
\begin{equation}
\hat \beta =  ({\bar H}'H)^{-1}{\bar{H}}'Y.
\end{equation}
Standard asymptotic arguments \citep[e.g.,][]{angrist2008mostly} give
\begin{equation}\label{sandwich}
 V_\beta \equiv \mr{var}(\hat\beta) = ({\bar H}'{\bar H})^{-1}{\bar H}'
 \mr{diag}\left[({H}\hat\beta - Y)^2\right]{\bar H}({\bar H}'{\bar H})^{-1}
 \end{equation} 
 as a consistent estimator for the sampling variance on $\hat\beta$.\footnote{The same variance can also be derived from Bayesian nonparametric arguments, e.g., as in \cite{taddy_heterogeneous_2015}.} 
 Note that (\ref{sandwich}) looks like the usual `sandwich' variance estimator \citep{white_heteroskedasticity-consistent_1980} for regression of $Y$ on ${\bar H}$, except that the sandwiched diagonal of squared residuals is calculated for the {\it treatment} input $H$
 rather than for the instruments $\bar H$.
 Finally, defining ${\eta}(x,p) =[1~\eta_{1}(x,p) \cdots \eta_{K}(x,p)]'$, the data-splitting estimate of our counterfactual prediction function $\cf(x,p)$ is 
 \begin{equation}
 \cfEst(x,p) = \hat\beta'{\eta}(x,p),
 \end{equation} with sampling variance 
 \begin{equation}\label{splitvar}
 \mr{var}\left(\cfEst(x,p)\right) = {\eta}'(x,p)~V_\beta~{\eta}(x,p).
 \end{equation}

This `post selection' variance might seem strange at first glance.  Our data splitting procedure views the first and second stage network fits as a search over different candidate models for a good set of basis functions and their conditional expectations. The variance in (\ref{splitvar})  summarizes frequentist uncertainty {\it conditional} on that model. 
However, since we are calculating uncertainty out-of-sample, there is a surprising amount of information in $V_\beta$.  For example, if the first stage does a poor job of modeling $F(p|x_i,z_i)$, then each $\bar\eta_{ik}$ will be far from $\eta(x_i,p_i)$ and the OLS residuals, $y_i - {\eta}_i'\hat\beta$, will be large (driving up the variance in $V_\beta$).  Moreover, Deep IV here can be viewed as automating the process of basis specification that authors such as \cite{newey2003instrumental,BCK2007}, and \cite{chenpouzo12} execute by hand and treat as given.

\subsection{Dropout variational Bayesian inference}

For unconditional uncertainty quantification, we look to
variational Bayesian inference.   For each of our two networks, we  fit a
variational distribution that approximates the posterior distribution over
network weights. In particular, we show that a network fit under dropout
\citep{srivastava2014dropout} parametrizes just such a variational
distribution and thus approximate Bayesian inference requires no more work
than dropout training (which is a good idea in any case).

We  introduce the procedure in terms of a generic neural network before describing the extension to Deep IV.   For simplicity we will ignore the network bias terms (i.e., the intercepts added to each layer input) in our description, but typically dropout is also applied to these terms. Define the $K_l \times K_{l-1}$ weight matrix from layer $l$ as $W_l$, where $K_{l}$ denotes the number of nodes in layer $l$.  In dropout training, this matrix is parametrized as $W_l = \Xi_l\Omega_l $, where $\Omega_l$ is a fixed $K_l \times K_{l-1}$  matrix and  $\Xi_l = \mr{diag}\left([\xi_{l1} \ldots \xi_{lK_{l}}]\right)$  with each $\xi_{kl}$ an independent  $\mr{Bernoulli}(c)$ random variable where $\ds{E}\xi_{kl} = c$.  During training, each SGD update for $W_l$ follows gradients with respect to $\Omega_l$ conditional upon a random realization for $\Xi_l$; {\it rows} of $\Omega_l$ corresponding to zero draws of $\xi_{lk}$ are thus `dropped' from the gradient calculations and updates.

 Variational Bayesian inference \citep[VB; e.g.,][]{bishop_pattern_2006} for parameters $W$ optimizes a parametric {\it variational distribution}, say $q(W)$, to be close to the true (but intractable) posterior  density $\mr{p}(W | \mc{D})$.  In particular, VB fits $q$ to minimize the Kullback-Leibler divergence $\ds{E}_q\left[\log q(W) - \log \mr{p}(W | \mc{D})\right]$, which is equivalent to solving
\begin{equation}\label{kldiv}
 \min_q \ds{E}_q \left[~\log q(W) - \log \mr{p}(\mc{D}|W) -  \log \mr{p}(W)~\right]
 \end{equation}  
where $\mr{p}(\mc{D}|W)$ is the likelihood and $\mr{p}(W)$ is the prior.  
 Say $W = \{W_0 \ldots W_L\}$ is the full set of network weights across layers, and similarly for $\Omega$ and $\Xi$. Then we define our variational $q(W)$ as the distribution over weights induced by setting $W_l = \Xi_l\Omega_l$ with $\xi_{lk} \stackrel{iid}{\sim} \mr{Bernoulli}(c)$.  Writing $W_{lk}$ for the $k^{th}$ column of $W_l$ and similarly for $\Omega_{lk}$, 
 \begin{equation}\label{vardist}
 q(W; ~c, \Omega)  =   
 \prod_{l = 1}^L \prod_{k = 1}^{K_{l-1}} \left(\ds{1}_{[W_{lk}=\Omega_{lk}]}c + \ds{1}_{[W_{lk}=0]}(1-c)\right).
 \end{equation}
 This might seem a strange variational distribution: the sole source of stochasticity is introduced by Bernoulli multipliers rather than the more common, e.g., Gaussian noise.  However it is a valid parametric distribution and, we shall see, it serves as a convenient posterior approximation.

If we treat $c$ as fixed in advance (or rather fixed after selection via cross validation; see below), then the first term in (\ref{kldiv}) is a {\it constant} under 
(\ref{vardist}).   Because each $\xi_{lk}$ is independent,
\begin{equation}\label{negent}
\ds{E}_q \log q(W) = 
\sum_{l = 1}^L \sum_{k = 1}^{K_{l-1}} \ds{E}_q \log q(W_{lk}) 
= \sum_{l = 1}^L K_{l-1} \left[c\log(c)
	+ (1-c)\log(1-c)\right],
\end{equation}
and this term is unchanging with $\Omega$.  
Removing this constant, and placing independent Gaussian priors $\mr{N}(w_{lkj}; 0, \lambda^{-1})$ on each scalar network weight, the VB minimization of (\ref{kldiv}) simplifies as 
\begin{equation}\label{qmap}
 \min_{\Omega} \ds{E}_q\left[ \mathscr{l}(\mc{D}|W) + \sum_{l = 1}^L \lambda\| W_l \|^2 ~\right]
 ~~=~~\min_{\Omega} \bigg\{~~\ds{E}_q \mathscr{l}(\mc{D}|W) +  c\lambda\| \Omega \|^2 ~~\bigg\}
 \end{equation}  
 where $\mathscr{l}(\mc{D}|W)$ is the negative log likelihood 
 (evaluated on the response given network output) and $\| \cdot \|$ denotes an entry-wise $\ell_2$ norm (i.e., ridge regularization).
Unbiased stochastic gradients for the objective in  $(\ref{qmap})$  against $\Omega$, the free parameters of our variational distribution, are approximated by taking gradients of $\mathscr{l}(\mc{D}|W)$ under  random realizations of $\Xi$ and with $\ell_2$ penalty weight $c\lambda$.  This is {\it exactly} how the gradients are calculated in dropout SGD, and thus when $c$ is treated as fixed {\it dropout training is variational inference under 
$q(W)$ as defined in (\ref{vardist}).}

\cite{gal2015dropout} make a similar point about dropout being interpretable as variational Bayesian inference.  However, they make this connection by introducing an additional model -- a  deep Gaussian process (GP) -- and arguing that dropout  neural network training approximates  the posterior distribution of a Deep GP after marginalizing out nuisance parameters. As our derivation above makes clear, there is no need to introduce the GP construction.\footnote{\cite{Gal2016Uncertainty} expands on the ideas of \cite{gal2015dropout}.   In both papers, the likelihood term $\ds{E}_q\left[ -\mathscr{l}(\mc{D}|W)\right]$ is viewed as the loss component that involves MC integration via dropout.  They derive  analytic relationships between the remaining terms in (\ref{kldiv}), $\ds{E}_q\left[\log q(W) - \log \mr{p}(W)\right]$, and the $\ell_2$ penalty applied on $\Omega$ (our notation) outside of dropout.  In contrast, we view dropout as MC integration for $\ds{E}_q\left[ -\mathscr{l}(\mc{D}|W) - \log \mr{p}(W)\right]$, which leads to the re-scaled $\ell_2$ penalty in (\ref{qmap}),  and show that $\ds{E}_q\left[\log q(W)\right]$ is fixed for a given $c$ and can be ignored.}

The results here show that dropout VB uncertainty is effectively {\it fixed in
advance} through the inverse dropout probability, $c$.  This should caution
users about being overly confident in the accuracy of dropout VB.  On the
other hand, one could treat $c$ as an unknown variational parameter. Write $K_{1:L} = \sum_{l = 1}^L K_{l-1}$ for the total number of outputs across all layers, and $\mr{ent}(c) = -\left[c\log(c)
	+ (1-c)\log(1-c)\right]$ as the entropy for a $\mr{Bernoulli}(c)$ random variable.  Then the full KL divergence minimization objective, over both $c$  and $\Omega$, is 
\begin{equation}\label{klwithc}
\ds{E}_{q(c,\Omega)} \mathscr{l}(\mc{D}|W) +  c\lambda\| \Omega \|^2 
- K_{1:L}\mr{ent}\left(c\right).
 \end{equation}  
The last term here, $K_{1:L}\mr{ent}\left(c\right)=
-\ds{E}_q\log q$ from (\ref{negent}), provides a {\it penalty} on the amount of dropout -- on
the choice of $c \in [0.5,1)$.  This penalty is minimized (lowering the KL)
for $c \approx 0.5$, a high dropout rate that makes it tougher to minimize the
other terms in (\ref{kldiv}).  At the same time, the negative log likelihood
can be made smaller under larger $c$ (less dropout), but
then the penalty term on $c$ grows.  In preliminary experimentation, we find  the $c$-values that minimize out-of-sample error also minimize the in-sample KL divergence of (\ref{klwithc}).  Thus, use of a test sample to tune dropout rates can be interpreted as approximate optimization for this variational parameter.  Since $c$ fully determines the amount of posterior uncertainty, such tuning is essential.

Application of dropout VB in Deep IV is straightforward.  The first stage network, $F_\phi$, just needs to be trained using dropout.  Then, in the second stage, each observation loss
$\left(y - \int \cf_\theta(p,x)dF_\phi(p|x,z)\right)^2$ is minimized while marginalizing over the joint dropout  variational distribution  $q(\theta,\phi) = q(\phi)q(\theta)$.  This is achieved by applying dropout uncertainty to draw a  single posterior realization of $\phi$ for each gradient calculation on $\cf_\theta$ while also using dropout in updates to $\theta$.
Since the KL expectation with respect to $q(\theta,\phi)$ is over the full loss function, you should  draw a single $\phi$ realization for each gradient update to $\theta$ (i.e., you don't want independent sampling here like in MC SGD). Note also that we are now interpreting the second stage loss as a negative log likelihood; this is in contrast to our frequentist inference where it remains simply a loss function.

\section{Simulation experiment}
\label{sec:sim}

\begin{figure}
\includegraphics[width=\textwidth]{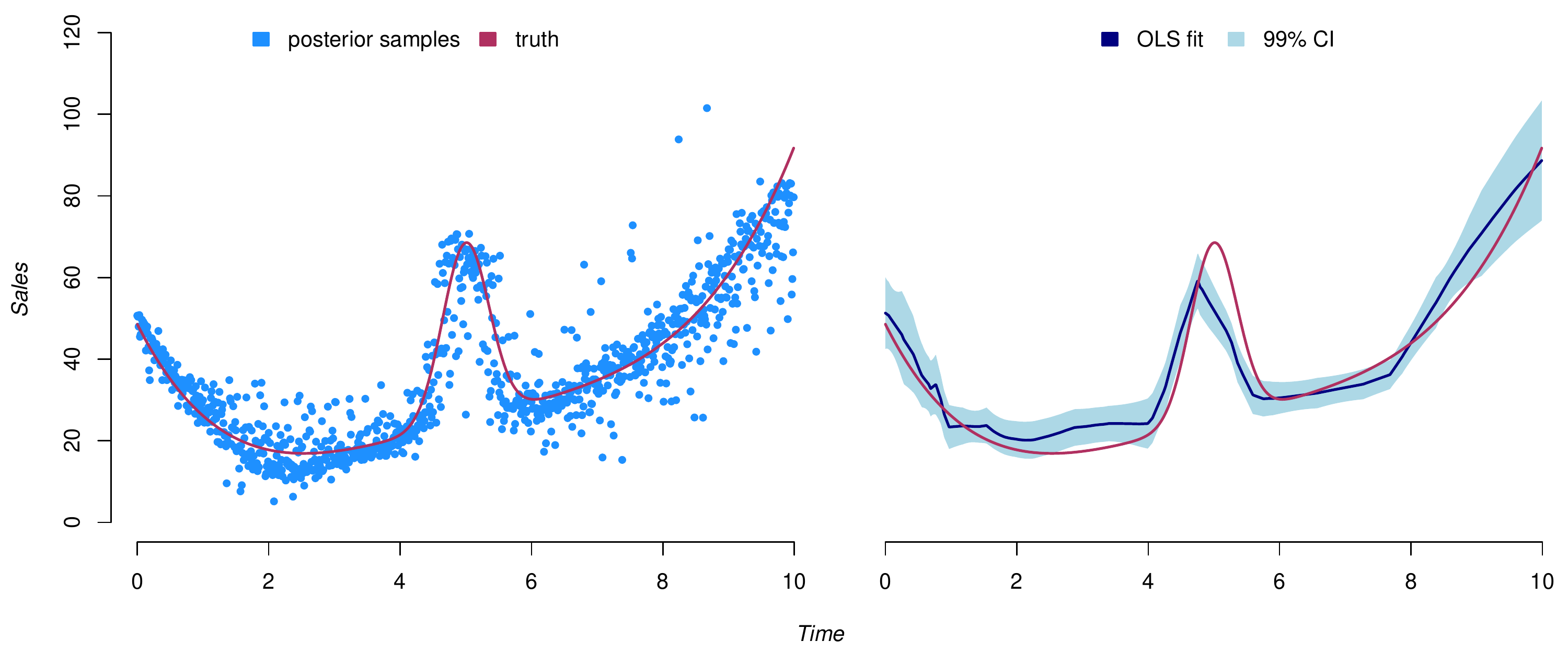}
\caption{\label{fig:storeinf} Bayesian (left) and Frequentist (right) inference for a central slice of the counterfactual function, taken at the average price and in our $4^{th}$ customer category. Since the price effect for a given customer at a specific time is  constant  in (\ref{simdgp}), the curves here are a rescaling of the customer {\it price sensitivity} function. }
\end{figure}

We illustrate in the context of a simple simulated economy. 
The  experiment is motivated by a story of customers with varying price sensitivity making consumption choices throughout the day, where prices are chosen strategically by the seller to move  with  average price sensitivity.  In addition to price $p$ -- our  policy variable -- the exogenous covariates are time $t \sim [0,10]$ and customer segment effect $s \in \{1, \ldots, 7\}$.  Time is uniformly distributed over its domain and, independently, customer segments are even-probability multinomial draws.  Sales, $y$, are then generated as
\begin{align}\label{simdgp}
 y &= 100 + s\psi_t + (\psi_t-2)p + e,\\
 p & = 25 + (z+3)\psi_t +  v \notag\\
& z, ~v  \sim \mr{N}(0, 1) ~~\text{and}~~e \sim \mr{N}(\rho v, 1-\rho^2),\notag
\end{align}
where $\psi_t$ is a (negative) nonlinear function of time that influences prices, demand, and price sensitivity.  The full expression is $\psi_t = 2\left((t - 5)^4/600  + \exp\left[-4(t - 5)^2\right] + t/10 - 2\right)$ and the true sales curves in Figure \ref{fig:storeinf} are shifted and scaled versions of this function.
The experiment was designed to allow us to  vary the endogeneity with a single parameter, $\rho\in[0,1]$, that spans an independent errors regime at $\rho=0$ to perfect correlation between $e$ and $v$ at $\rho=1$.

\begin{figure}
\includegraphics[width=\textwidth]{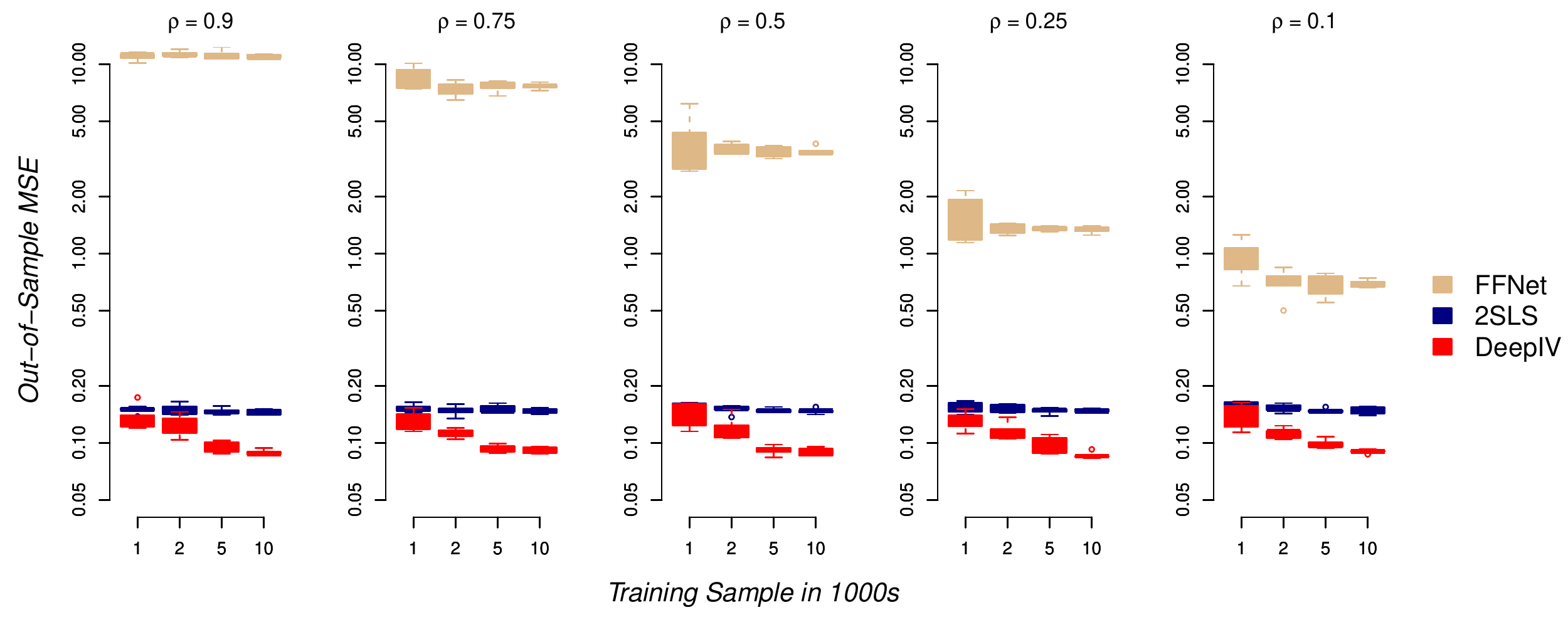}
\caption{\label{fig:storeperf} Out-of-sample predictive performance for different levels of endogeneity ($\rho$).  Note that the test samples are generated with independent errors conditional upon a fixed grid of price values.  This breaks the endogeneity that exists in the training sample, which is why the feed forward network does so poorly. }
\end{figure}

Our target counterfactual function is $\cf(t,s,p) = s\psi_t + (\psi_t-2)p$.  In out-of-sample evaluation, we compare estimated $\cfEst$ against the truth evaluated over a {\it fixed grid} of price values (with $[t,s]$ sampled as in the original DGP).  This breaks
the endogeneity that exists in the training sample, so that our out-of-sample errors are representative of the structural error that is of interest for counterfactual inference.  In addition to Deep IV, we consider  a regular feed-forward network (FFNet) and standard two-stage least squares (2SLS).  \footnote{Both Deep IV and FFNet have a single hidden layer of 50 nodes, and we apply a dropout rate of 0.5 for Deep IV. }  We evaluate structural mean square error (MSE) while varying both the number of training examples and the amount of endogeneity. The results are summarized in Figure \ref{fig:storeperf}. 
Both 2SLS and our Deep IV model are designed to solve the endogeneity problem, and we see that their performance is mostly unaffected by changes in the amount of endogeneity.  2SLS is constrained by its homogeneity and linearity assumptions, so that it does not improve with increasing data. However, it still does much better than FFNet.  This naive ML  is doing a {\it good} job of estimating $\cf(t,s,p) + \ds{E}[e|p]$, but a terrible job of recovering the true counterfactual.  As endogeneity drops, so that $\ds{E}[e|p]$  decreases, FFNet improves. However even at low levels of endogeneity it remains far worse than simple 2SLS.  In contrast,  Deep IV  is the best performing model throughout and its performance improves as the amount of data grows.

Figure \ref{fig:storeinf} uses the Bayesian and Frequentist techniques from Section \ref{sec:inference} to obtain uncertainty estimates for our model's predicted sales over time.  This is a slice of $\cf$ evaluated at averages $p=25$ and $s=4$, trained on a large set of 1 million observations.\footnote{For this network, we have four layers with widths of 256,128,64, and 32.  The test-sample MSE at $c=.99$ is 0.026 using the same criterion as in Figure \ref{fig:storeperf}.  Thus the error rate continues to drop with additional data in this example, so long as the network is allowed to grow in complexity. } Since expected sales for a given customer at a specific time are linear in price, the curves here are a rescaling of the customer {\it price sensitivity} function.  This is the object that we need to recover for structural counterfactual inference.  The Bayesian inference here is for an inverse dropout rate of $c=0.99$, which was tuned to be optimal in out-of-sample prediction.  We see that DeepIV is able to mostly recover the true counterfactual shape.  The Bayesian procedure provides wider uncertainty than the conditional inference obtained through data splitting, but this is completely determined by our choice of $c$.   We have no strong argument for one inference  procedure over the other at this time.



\section{Discussion}
\label{sec:disc}

The next generation of problems in ML involve moving from raw prediction tasks into more complex decision-making domains.  In addition, we want our AI solutions to be transparent and to have the ability to respect notions of, e.g., fairness. All of these needs require a knowledge of the true structure of the processes that we are modeling and, hence, causal inference.  The work in this paper is a step in the direction of Causal AI.  The Deep IV framework shows that it is possible to take generic ML procedures, trained via SGD, and to use econometric theory to stack them together into a system that provides reliable answers to causal questions.  This is a recipe for Artificial Economic Intelligence, and we think that much more can come from such an approach.

\bibliography{deepiv}
\end{document}